\documentclass[conference]{IEEEtran}
\IEEEoverridecommandlockouts
\usepackage{cite}
\usepackage{amsmath,amssymb,amsfonts,mathrsfs,amsthm}
\usepackage{algorithm}
\usepackage{algpseudocode}
\usepackage{comment}
\usepackage{graphicx}
\usepackage{textcomp}
\usepackage{xcolor}
\usepackage{url}
\usepackage{adjustbox}
\usepackage{balance}
\usepackage{subcaption}
\usepackage{makecell}
\usepackage{orcidlink}
\usepackage{stfloats} 

\PassOptionsToPackage{bookmarks=false}{hyperref}

\def\BibTeX{{\rm B\kern-.05em{\sc i\kern-.025em b}\kern-.08em
    T\kern-.1667em\lower.7ex\hbox{E}\kern-.125emX}}
\begin{document}

\title{Design and Performance Evaluation of a BLE-Based IoT Authentication System\\

}

\author{\IEEEauthorblockN{Nitesh Yadav, Vashisht Kumar, and Sachin Kadam}
\IEEEauthorblockA{{Electronics and Communication Engineering Department} \\
{Motilal Nehru National Institute of Technology Allahabad, Prayagraj, UP 211004, India}\\
Email: nitesh.2024sp13@mnnit.ac.in, vashisht.2024sp21@mnnit.ac.in, sachink@mnnit.ac.in}
}

\maketitle

\begin{abstract}
Bluetooth Low Energy (BLE) is widely used in modern IoT systems because it consumes very little power, saves energy, and allows for simple device connectivity; however, maintaining security and communication reliability remains a challenge. In this paper, an authentication system is designed using industry-grade BLE-enabled nodes (nRF5340 development kit), that include a peripheral node with a keypad for entering a PIN and a central node with an LCD display. The entered PIN is sent wirelessly from the peripheral node to the central node via BLE technology, where it is verified in real time and displayed as correct or incorrect. Next, only after successful authentication can the peripheral node send data to the central node. In addition to authentication, the peripheral node can measure temperature in real time using the temperature sensor interfaced to it and send it wirelessly to the central node, where it can be displayed on the LCD interface. Received Signal Strength Indicator (RSSI) values are collected during experiments under various scenarios to evaluate the system's performance. We see that the signal strength (measured in terms of RSSI values) is strong at close range but weak as distance increases, indicating a decaying logarithmic pattern. The system also has low latency, which allows for quick input and output, and it uses PIN-based authentication to ensure security and prevent misuse. The entire system seamlessly integrates communication, sensing, and security, making it suitable for smart access control and wireless monitoring systems, including home automation.
\end{abstract}

\section{Introduction}
Wireless communication is vital for deploying Internet of Things (IoT) systems in various applications, such as smart home environments, small-scale industrial setups, and basic remote monitoring configurations~\cite{sah2025comprehensive}. Bluetooth Low Energy (BLE) is a popular short-range communication protocol due to its low power consumption and compatibility with resource-constrained embedded platforms~\cite{koulouras2025evolution}. Despite these advantages, communication capability alone is insufficient for many practical deployments, as access control applications, in particular, require some form of user authentication to prevent unauthorized entry; the lack of such a mechanism leaves the system vulnerable to misuse. With this dual requirement in mind, our work proposes a BLE-based system designed to provide PIN-based authentication followed by wireless data transfer, with a matrix keypad serving as the input interface through which the user enters a numeric PIN, which is then transmitted wirelessly to a second device, verified against a stored reference value, and the result—access granted or denied—displayed on a connected LCD screen. In addition, a temperature sensor is integrated into a device, with readings continuously transmitted via BLE and displayed on the same display, broadening the system's scope beyond access control to include basic environmental monitoring. 

\begin{figure}[htbp]
\centerline{\includegraphics[scale=0.2]{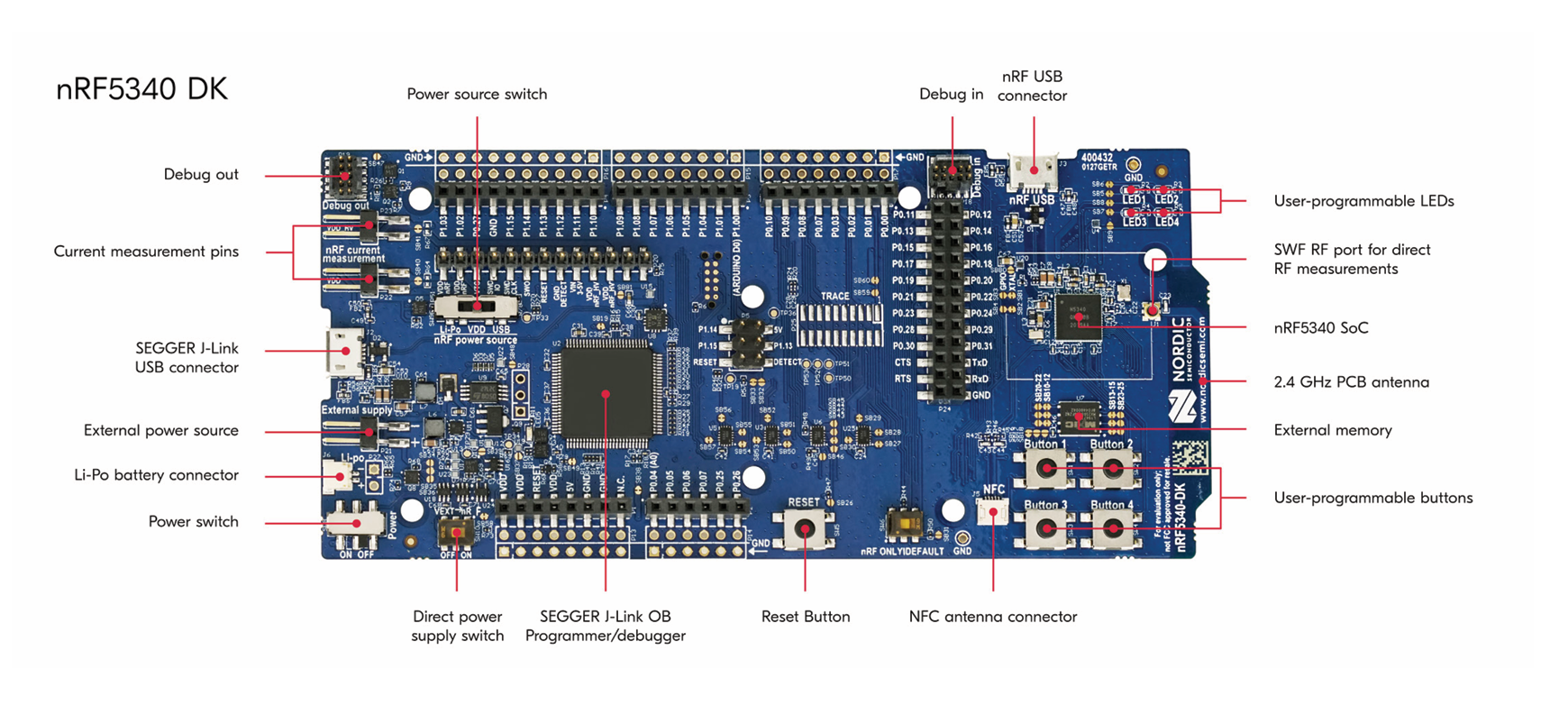}}
\caption{nRF5340 development kit (DK) hardware platform~\cite{nordic2021productbrief}.}
\label{fig:nRF5340}
\end{figure}

The performance in terms of data transfer is evaluated by recording RSSI values over varying distances under various conditions, and it is discovered that closer proximity produces stronger and more stable signals, whereas increasing separation causes a gradual decline in signal strength as well as occasional irregularities, which are worsened by physical obstructions such as interior walls, trees, and so on. These findings reflect the practical constraints of BLE in real-world settings and contribute to a more thorough understanding of its behavior. Overall, the proposed system provides a small and simple solution that integrates communication and basic security into a single BLE framework, with potential applications including smart door lock systems, home automation, and small-scale wireless monitoring setups. Figure~\ref{fig:nRF5340} depicts the hardware platform for the nRF5340 development kit (DK).

\section{LITERATURE REVIEW}\label{Sec:Literature}
Bluetooth Low Energy (BLE) is a prevalent IoT system due to its minimal power usage and simplicity. Its performance in the context of latency, throughput, energy efficiency, and reliability has been investigated by many researchers. The authors in~\cite{baghel2024latency} studied the impact of advertising interval on the latency of data reception and demonstrated that the inappropriate choice of parameters might raise delay and collisions in BLE communication. This implies that real-time applications require parameter tuning. The authors in~\cite{gautam2025throughput} explored the maximum throughput that can be achieved in BLE systems and discovered that larger payload sizes may enhance the efficiency, but retransmissions and bit errors may decrease the overall performance. This demonstrates that there is a trade-off in the data size and reliability. Another significant aspect of the IoT systems is energy efficiency. A subsequent study by~\cite{baghel2025livestock} suggests a BLE-based sensing system that consumes less power. It also demonstrated that the battery life can be increased several times by minimizing unnecessary transmissions. This can be handy in long deployments where devices are subject to limited power. The introduction of BLE 5.x has enhanced the ability to transmit data with the added features such as extended advertising. The work in~\cite{gautam2025energy} shows that a large amount of sensor data can be transmitted with low packet loss, even in dense networks. However, the lack of acknowledgement in advertising-based communication still affects reliability. The communication range is also an important part of BLE systems. Next, the work in~\cite{verma2025rpm} demonstrates that using LE Coded Physical Layer (PHY) can extend communication range and can reduce packet loss, but it could also reduce data rate. This highlights the need to balance range, power consumption, and performance. RSSI-based analysis has been commonly used to evaluate signal strength and estimate the distance between devices. Studies such as~\cite{ramirez2021rssi} and~\cite{biju2022rssi} showed that RSSI varies with distance and environmental conditions. However, RSSI values are affected by noise and interference, and therefore filtering techniques are very often used to improve accuracy~\cite{mittal2021localization}. Security has been another key requirement in BLE-based IoT systems. 
According to~\cite{hussain2018secure}, a BLE-based system can be vulnerable to unauthorized access without proper authentication mechanisms in place. This demonstrates the importance of user authentication. RSSI-based techniques, combined with data-driven approaches, are used in environmental monitoring and prediction systems~\cite{keshavarz2025soil}. This suggests that BLE has a high potential for use in future intelligent IoT applications. In~\cite{khanchuea2019iot}, a multi-protocol IoT gateway was designed for smart home and building automation, where BLE, WiFi, and ZigBee are used. The goal is to use multi-hop communication to connect various types of devices in a single system and increase coverage. This increases the flexibility of communication, but the system becomes more complex, and problems such as interference may arise.
In~\cite{devi2025blockchain}, a blockchain-based method is used to make BLE pairing more secure. It aids in the reduction of issues such as unauthorized access and the prevention of Man-in-the-Middle attacks. In~\cite{chandan2018ble}, practical testing demonstrates that a BLE-based device can be accessed without proper security. This clearly demonstrates that many systems do not have adequate protection. According to~\cite{uher2016sleep}, a denial of sleep attack causes devices to remain active and unable to enter low power mode, resulting in faster battery drain. In~\cite{zhang2023fingerprint}, a method is used to identify BLE devices using packet information, thereby separating genuine and unknown devices. This can help to improve overall system security.
In~\cite{rattal2025ai}, AI-based methods are used to improve energy efficiency and performance, especially when multiple devices are connected. In~\cite{marques2025doorlock}, a BLE-based smart door lock system is developed, demonstrating how BLE can be used in real applications such as access control. Several studies have looked at BLE from various angles, including security, localization, and general IoT communication. In~\cite{ray2018testing}, a testing framework was proposed to identify security vulnerabilities in BLE devices before deployment. Wireless key sharing was also explored as a practical feature for everyday use, but concerns about power efficiency and overall reliability still remain. 
In attendance monitoring, the work in~\cite{chavhan2025attendance} used a smartphone as a beacon combined with biometric authentication to prevent proxy attendance. It was observed that the BLE signal strength weakens noticeably when walls or other obstacles are present. In indoor localization, the work in~\cite{das2018microlocation} used path-loss models and curve fitting before triangulation to improve distance estimation, while the authors in~\cite{kalbandhe2016ips} used signal strength and transmission power with BLE tags as a low-cost alternative to Wi-Fi-based positioning. In~\cite{kashani2024dataset}, machine learning was introduced through a BLE signal dataset designed for Wireless Body Area Networks. Similarly, the work in~\cite{jeon2018ble_survey} provided a broader survey of BLE beacon applications in smart environments, highlighting signal fluctuation as a persistent challenge. In~\cite{gilmartinez2026vehicle}, vehicle tracking was demonstrated using BLE and RSSI combined with antenna techniques. The authors in~\cite{manzoor2018relay} proposed using smartphones as relay nodes to extend the communication range of low-power sensors. In~\cite{shah2025beamforming}, RSSI was used to select the best relay node, improving reliability. Similarly, in~\cite{manzoor2018relay}, beamforming antenna techniques were integrated with BLE for healthcare environments with high signal attenuation.  The work in~\cite{harris2016dense_full} opportunistic listening and aggregation to reduce collisions in dense IoT deployments. The authors in~\cite{boualouache2015ble} presented a simple smartphone-based data collection system suitable for short-range IoT use, while the work in~\cite{gupta2025localization} developed an ESP32-based indoor localization system using RSSI and machine learning, noting that environmental factors such as walls and interference significantly affect accuracy. The work in DL-based BLE localization~\cite{11352866} uses RSSI as a main input to a machine learning model to estimate position without focusing on the signal's physical trends. In a BLE attendance system~\cite{11473285}, UUIDs are used to identify devices, with the primary goal of attendance marking rather than signal analysis. The path loss study in~\cite{kareem2025empirical} uses a log-distance model to explain how RSSI changes with distance and provides insight into signal attenuation in various conditions. In comparison, this work investigates RSSI behavior in various environments and applies it to signal variation-based authentication.

Overall, these studies confirm that BLE is well suited for low-power IoT applications; however, environmental conditions and signal variability remain practical challenges that must be carefully considered during system design. Although generic BLE-enabled nodes make the system inexpensive and simple to build, they are best suited for academic or experimental applications. In contrast, the proposed work employs the nRF5340 DK, a more advanced and industry-standard platform. It offers better performance, more reliable communication, and improved BLE features, making it more suitable for real-world use.

\section{SYSTEM MODEL}

\subsection{System Overview}
\begin{figure*}
\centerline{\includegraphics[scale=0.42]{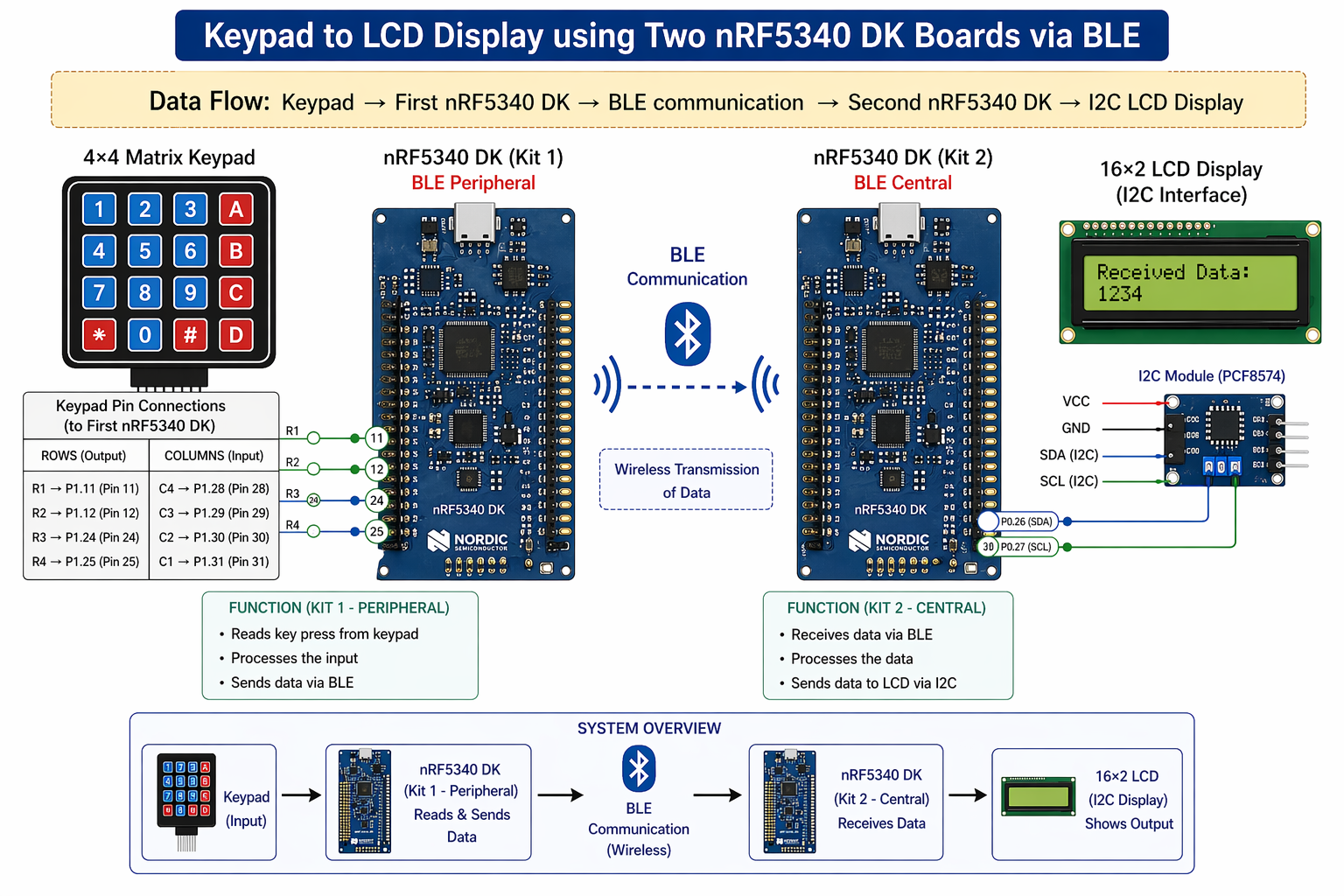}}
\caption{Our proposed system model shows two nRF5340 DK modules, with one (peripheral node) interfaced to a keypad to get user input and another (central node) interfaced to an LCD display to observe the user-provided input. The data transfer between the modules is using BLE technoogy.}
\label{fig:SystemModel}
\end{figure*}
The proposed system is a Bluetooth Low Energy (BLE) configuration that ensures effective real-time communication with authentication, as seen in Fig.~\ref{fig:SystemModel}. The transmitter in this system is the peripheral node denoted as BLE$_p$, and the receiver is the central node denoted as BLE$_c$. A keypad is interfaced with BLE$_p$ that enables the user to enter the PIN, and an LCD is interfaced to BLE$_c$ that displays the output. Using the keypad, the user enters a PIN at BLE$_p$, which is wirelessly sent via BLE to the BLE$_c$. After receiving the PIN, BLE$_c$ checks whether the entered PIN is correct or not, which is displayed on the LCD. This approach makes it suitable for secure access and authentication applications. Once the authentication is successful, the BLE$_p$ can transfer its data wirelessly to the BLE$_c$ using BLE technology. If the authentication is not successful, then the LCD interfaced to BLE$_c$ displays `Wrong PIN, enter again.' And a counter that counts the successive wrong PIN entries is incremented. If this counter reaches the `MaxCount' value, then the opportunity to access BLE$_p$ is blocked for a certain duration.  Both the nodes work swiftly together to provide a response in no time at all for a satisfactory user experience. This system effectively integrates wireless communication, authentication, and sensing  the data in a straightforward and practical manner, making it applicable to smart locks, access control systems, and basic sensing systems.
\subsection{Hardware Components} \label{Sec:Hardware}
The system proposed is designed using a basic set of hardware components to transmit and receive messages from users and display the messages. Nodes employed in the setup are the nRF5340 development kits (DKs), which support data transfer using BLE. Two DKs are used, where one acts as BLE$_p$ and the other one acts as BLE$_c$. To allow the user to enter the pin, BLE$_p$ is connected to a 4×4 matrix keypad as shown in Fig.~\ref{fig:SystemModel}. Each time a key is pressed, the microcontroller of BLE$_p$ reads the data and prepares it for transmission over BLE. The BLE$_c$ contains a 16×2 LCD connected through an I2C interface that is used to display the received data and the result display of the PIN verification process. Thus, the output is clearly visible on the receiver side. In addition to the authentication process mentioned above, a temperature sensor is connected to BLE$_p$ to demonstrate IoT-based data transmission as the BLE$_c$ collects the  data from the temperature sensor connected to it. Subsequently, after the successful authentication, data from BLE$_p$ is sent to the BLE$_c$ via BLE and displayed on the LCD screen. These components are integrated to facilitate seamless messaging that responds in real-time while operating reliably.

\subsection{Temperature Data Transmission}
In our proposed system, a temperature sensor measures the ambient temperature. The sensor is linked to the peripheral node, which retrieves data, such as temperature readings, from the sensor. This data is then sent to the central node using BLE technology. When the central node receives the temperature, it displays it on the LCD. This demonstrates that we have a solution that works well in basic real-time monitoring applications.

\subsection{RSSI Value Measurements}
The performance of the proposed system is evaluated by measuring the Received Signal Strength Indicator (RSSI) values with respect to distance in various kinds of environments. 
For this purpose, we place the nodes at varying distances apart and record the RSSI values at each location. 
We use these values to determine the BLE system's communication range and overall performance. 
To validate the experimental findings, we compare the measured RSSI values to the estimated analytical RSSI values.
The RSSI value at a given distance $d$ can be estimated using a log-normal propagation model, as~\cite{fabris2025aoa}:
\begin{equation}
{RSSI}(d) = {RSSI}(d_0) - 10\alpha \log_{10}\left(\frac{d}{d_0}\right) + X_{\sigma}
\label{eq:rssi_model}
\end{equation}
where RSSI(d) is the RSSI (in dBm) at distance $d$, RSSI$(d_0)$ is the RSSI at the reference distance $d_0$ (typically 1m), $\alpha$ is the path-loss coefficient, and $X_{\sigma} \sim \mathcal{N}(0,\sigma^2)$ models the shadowing effect (in dB). The comparison results of experimental and analytical RSSI values vs. distance under various categories is described in Section~\ref{Sec:RSSIvsDistance}.


\section{EXPERIMENTAL RESULTS AND ANALYSIS}

\subsection{Experimental Setup}


As discussed in Section~\ref{Sec:Hardware}, we use two nRF5340 development boards for our experiment, in which one acts as a BLE$_p$ node and the other acts as a BLE$_c$ node. We have interfaced a keypad to the BLE$_p$ and an LCD to the BLE$_c$. Also, we have a temperature sensor, which is connected to the BLE$_p$ for measuring the environment temperature. We use common interfaces like GPIO and I2C for component interaction, and we make sure both nodes are powered appropriately. In order to facilitate seamless communication and real-time operation, we chose a straightforward but well-organized setup. In order to ensure robust BLE communication, we also situate the devices within range and make an effort to ensure correct wiring in order to rule out any signal problems. To ensure that every component is operating correctly, the system is tested several times.

\begin{figure}
\centering
\begin{subfigure}{.24\textwidth}
\centering
\begin{adjustbox}{width = 0.95\columnwidth}
\includegraphics[width=0.99\textwidth]{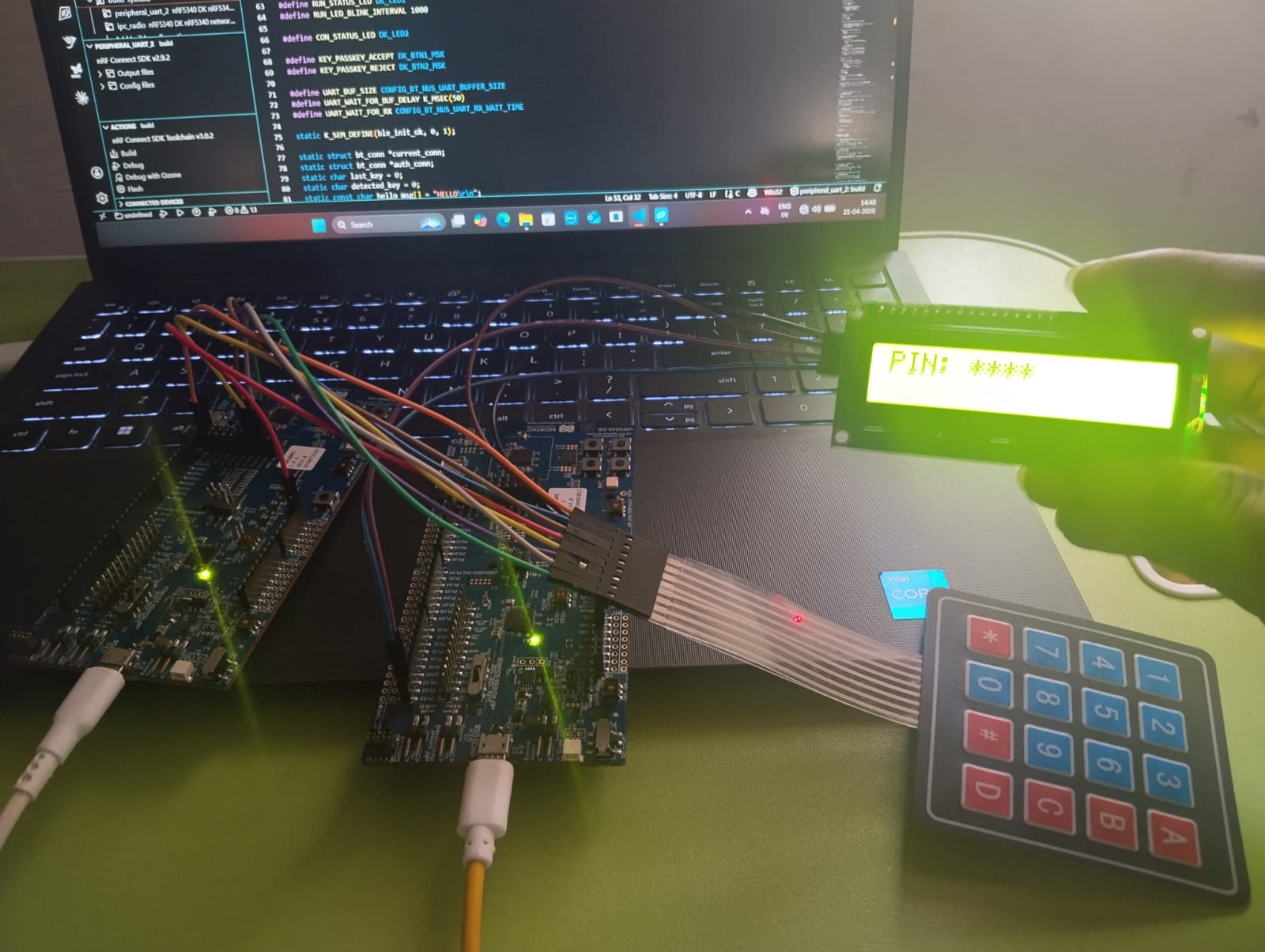}
\end{adjustbox}
\caption{}
\label{fig:PinEntry}
\end{subfigure}%
\begin{subfigure}{.24\textwidth}
\centering
\begin{adjustbox}{width = .95\columnwidth}
\includegraphics[width=0.99\textwidth]{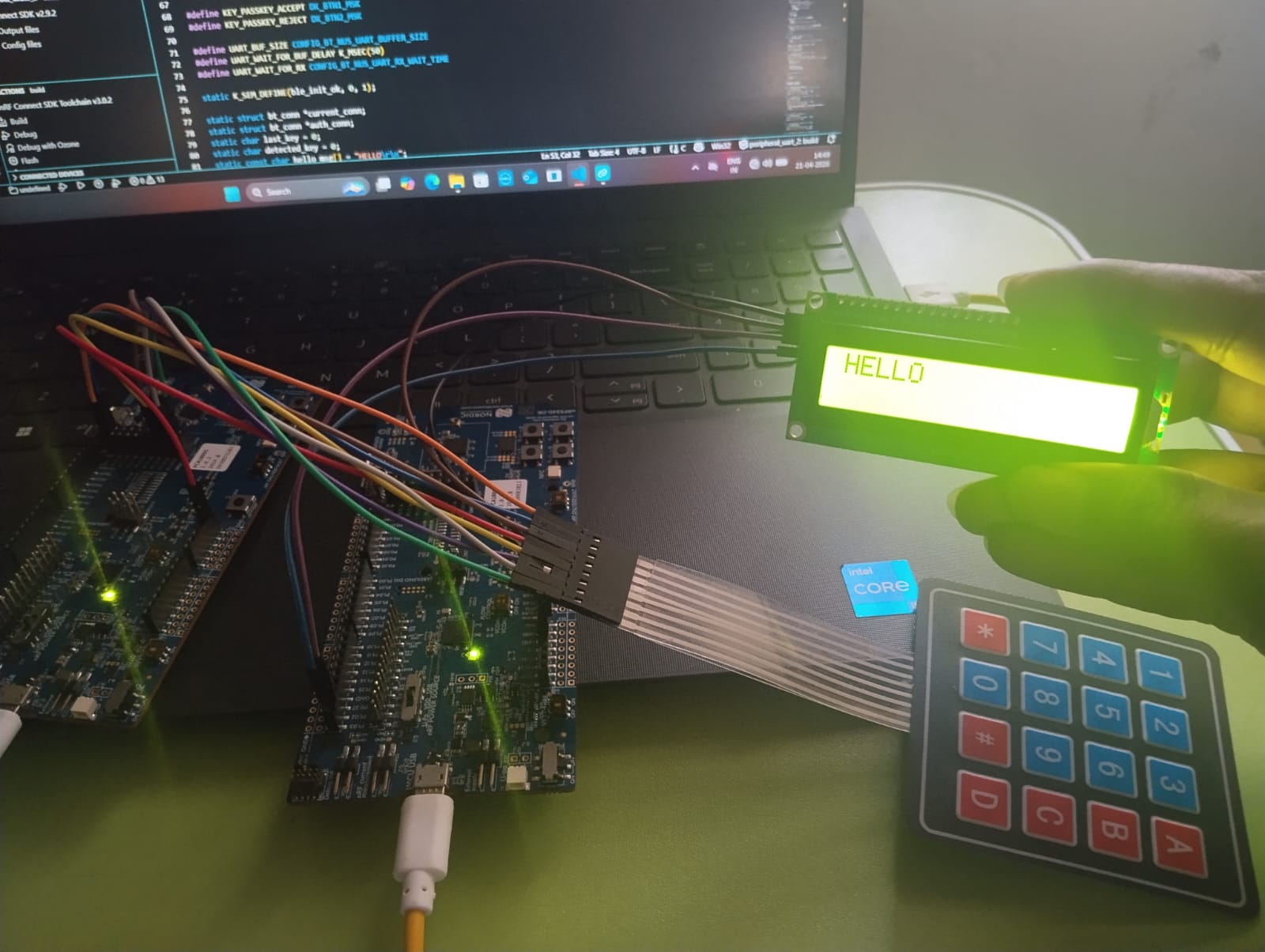}
\end{adjustbox}
\caption{}
\label{fig:CorrectPin}
\end{subfigure}
\caption{(a) Hardware setup of BLE authentication system (entering PIN at BLE$_p$ for access to BLE$_c$). (b) Successful authentication  (correct PIN entry leads to ``HELLO'' message display on the LCD screen).}
\label{fig:ExptAuthentication}
\end{figure}

Next, we perform two types of experiments, viz., authentication verification and RSSI vs. distance measurements.
\subsection{Authentication Verification}
The system is tested for PIN-based authentication, where both correct and incorrect PIN inputs are verified, as shown in Figs.~\ref{fig:ExptAuthentication}. The numbers $0-9$ and letters $A-F$ are used as characters for PIN entry, $*$ button is used for resetting, and $\#$ button is used for submitting the entered PIN to the BLE$_p$. Next, we set a 4-digit PIN for the BLE$_c$ node. To communicate with BLE$_c$, every BLE node first needs to authenticate by entering the correct PIN. The BLE$_p$ (left-side node in Figs.~\ref{fig:ExptAuthentication}) enters the PIN using a keypad interfaced to it. Every character typed on the keypad is transferred via BLE technology to the BLE$_c$. The BLE$_c$ converts these characters into $*$ marks and sends them to the LCD screen for display to the user, as shown in Fig.~\ref{fig:PinEntry}. If the PIN is wrong then `Wrong PIN, enter again' is displayed; otherwise, the `HELLO' message is displayed, as shown in Fig.~\ref{fig:CorrectPin}. Once the authentication is successful, the BLE$_p$ is able to transfer its data to the BLE$_c$. 

The performance of data transfer in terms of RSSI vs. distance measurements under various scenarios is discussed in the next subsection.

\subsection{RSSI vs Distance Analysis} \label{Sec:RSSIvsDistance}


The performance of the proposed BLE-based authentication system is measured in RSSI using two nRF5340 BLE-enabled development boards. We consider four scenarios: indoor (in Section~\ref{Sec:Indoor}), outdoor (in Section~\ref{Sec:Outdoor}), combined indoor and outdoor (in Section~\ref{Sec:InOutdoor}), and ground-level testing (in Section~\ref{Sec:Ground}). 
RSSI values are measured at varying distances to determine how signal strength varies with distance. In general, the signal strength weakens as the distance increases. When tested at the same distance in different environments, the RSSI value differs significantly. We can conclude from the measured RSSI values that signal strength is affected by both distance and surroundings. We also compared the experimentally measured RSSI values with the estimated RSSI values~\eqref{eq:rssi_model} and found that they closely match. Next, we look at how the proposed system performs in different scenarios.

\begin{figure}
\centering
\begin{subfigure}{.24\textwidth}
\centering
\begin{adjustbox}{width = 0.95\columnwidth}
\includegraphics[width=0.99\textwidth]{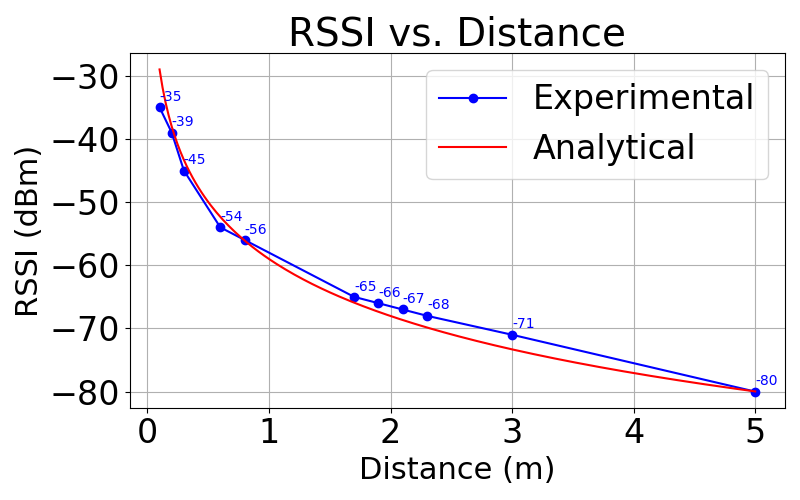}
\end{adjustbox}
\caption{}
\label{fig:Indoor_Sub}
\end{subfigure}%
\begin{subfigure}{.24\textwidth}
\centering
\begin{adjustbox}{width = .95\columnwidth}
\includegraphics[width=0.99\textwidth]{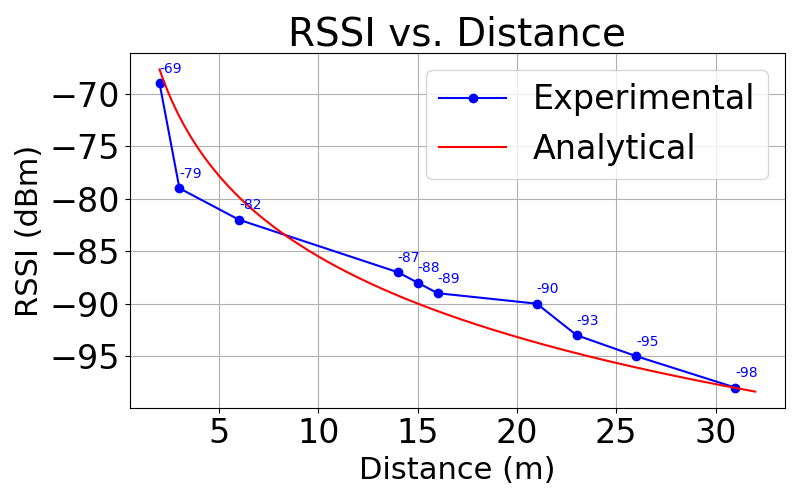}
\end{adjustbox}
\caption{}
\label{fig:Outdoor_Sub}
\end{subfigure}
\caption{RSSI values with respect to the varying distance in the indoor environment (a) and in the outdoor environment (b).}
\label{fig:IndoorOutdoor}
\end{figure}
\subsubsection{Indoor Environment} \label{Sec:Indoor}
In this scenario, both nodes, BLE$_p$ and BLE$_c$, are kept in a closed room, and RSSI values are measured at BLE$_c$ while keeping BLE$_p$ at different distances. The measured values are noted down and plotted in Fig.~\ref{fig:Indoor_Sub}. Additionally, the analytical RSSI values~\eqref{eq:rssi_model} with $\alpha = 3.1$ are plotted on the same graph for comparison purposes. The experimental RSSI values decrease logarithmically with distance, just like analytical RSSI values. The drop in RSSI values is steep at short distances, ranging from 0.1m to 0.6m. The mid-range (0.6m to 3m) exhibits stable behavior, while noise and multipath effects significantly reduce RSSI values at long ranges ($>$ 3m). 


\subsubsection{Outdoor Environment} \label{Sec:Outdoor}
In this scenario, both nodes, BLE$_p$ and BLE$_c$, are kept on our institute's outdoor sports ground, and RSSI values are measured at BLE$_c$ while stationing BLE$_p$ at varying distances. The measured values are plotted in Fig.~\ref{fig:Outdoor_Sub}. In addition, the analytical RSSI values~\eqref{eq:rssi_model} with $\alpha = 2.55$ are plotted on the same graph for comparison. The experimental RSSI values decrease logarithmically with distance, as do analytical RSSI values. In outdoor conditions, RSSI decreases as distance increases, but slower than in the indoor environment. The decaying slope is very steep at short distances (2-3m) before becoming very small. Mid-range (5-20m) and long-range ($>$20m) are extremely stable.  


\begin{figure}
\centering
\begin{subfigure}{.24\textwidth}
\centering
\begin{adjustbox}{width = 0.95\columnwidth}
\includegraphics[width=0.99\textwidth]{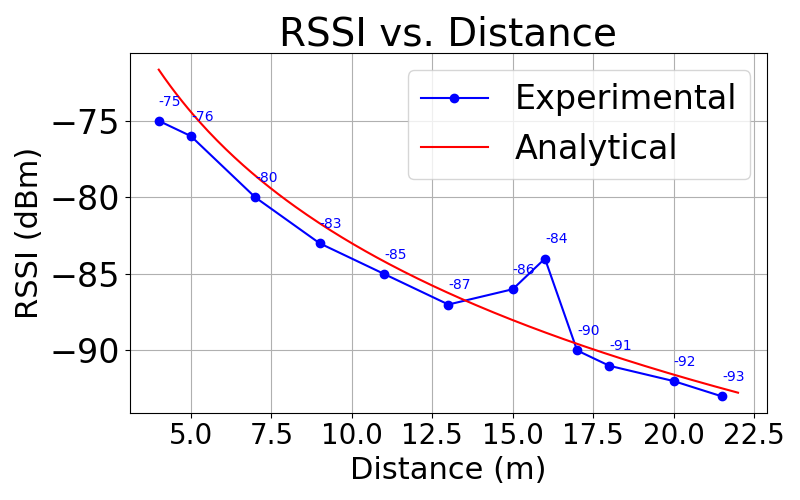}
\end{adjustbox}
\caption{}
\label{fig:InOutdoor_Sub}
\end{subfigure}%
\begin{subfigure}{.24\textwidth}
\centering
\begin{adjustbox}{width = .95\columnwidth}
\includegraphics[width=0.99\textwidth]{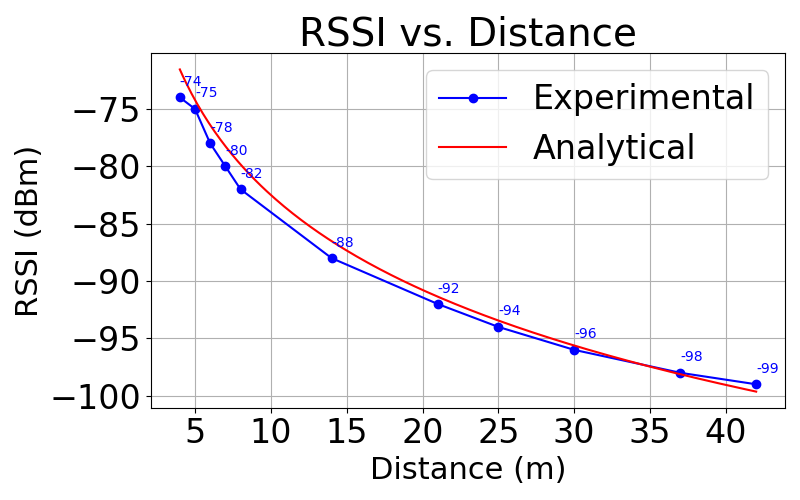}
\end{adjustbox}
\caption{}
\label{fig:Ground_Sub}
\end{subfigure}
\caption{RSSI values with respect to the varying distance in the combined indoor and outdoor environments (a) and at ground levels with obstructions (b).}
\label{fig:IndoorOutdoorGround}
\end{figure}
\subsubsection{Indoor and Outdoor Combined Environment} \label{Sec:InOutdoor}
In this scenario, one node, BLE$_c$, is kept in a closed room, while the other node, BLE$_p$, is placed outside the room. RSSI values are measured at BLE$_c$ while stationing BLE$_p$ at varying distances outside the room. Figure ~\ref{fig:InOutdoor_Sub} shows the measured values. In addition, the analytical RSSI values \eqref{eq:rssi_model} with $\alpha = 2.85$ are plotted on the same graph for comparison. The experimental RSSI values decrease logarithmically with distance, with the indoor region decaying faster and the outdoor region decaying slower, similar to the analytical RSSI values. 
Around 15.5-18 m, a noticeable change occurs, with the RSSI value slightly increasing due to the environmental change from an indoor to an outdoor scenario. Overall, the behavior falls somewhere between indoor and outdoor scenarios. 

\subsubsection{Ground Level Testing}\label{Sec:Ground}
In this scenario, both nodes, BLE$_p$ and BLE$_c$, are kept outdoors, but with several plants in between to block line of sight, and RSSI values are measured at BLE$_c$ while BLE$_p$ is placed at varying distances. Figure ~\ref{fig:Ground_Sub} depicts the measured values. Furthermore, the analytical RSSI values ~\eqref{eq:rssi_model} with $\alpha = 2.75$ are plotted on the same graph for comparison. Both the analytical and experimental RSSI values decrease logarithmically with distance. In this case, the decay is smoother in the long-range but faster in the short and medium-range distance values.   


Comparison of existing works with the proposed authentication system is shown in Table~\ref{Tab:AuthenticCompare}.
\begin{table}
\centering
\small
\setlength{\tabcolsep}{.4pt}
\vspace{+3mm}
\caption{Comparison of existing works with proposed BLE-based authentication system}
\begin{tabular}{|c|c|c|c|c|}
\hline
\textbf{Ref.} & \textbf{Security} & \textbf{Real-Time} & \textbf{Hardware} & \textbf{RSSI Use} \\
\hline
{\cite{chavhan2025attendance}} & Partial & Yes & Partial & Yes \\
\hline
{\cite{das2018microlocation}} & No & No & No & Yes \\
\hline
{\cite{kalbandhe2016ips}} & No & No & No & Yes \\
\hline
{\cite{kashani2024dataset}} & ML-based & No & No & Yes \\
\hline
{\cite{jeon2018ble_survey}} & No & No & No & No \\
\hline
{\cite{gilmartinez2026vehicle}} & No & Yes & Partial & Yes \\
\hline
{\cite{manzoor2018relay}} & Partial & Yes & Partial & Yes \\
\hline
\textbf{This} & \textbf{Yes} & \textbf{Yes} & \textbf{Yes} & \textbf{Yes} \\
\textbf{ Work } & \textbf{ (PIN-based) } &  & \textbf{ (nRF5340 DK) } &  \\
\hline
\end{tabular}
\label{Tab:AuthenticCompare}
\end{table}
The proposed work utilizes an industry-grade BLE platform (nRF5340) and provides a physically interpretable RSSI-based analysis with authentication capability, whereas most of the existing works shown in Section~\ref{Sec:Literature} and Table~\ref{Tab:AuthenticCompare} primarily rely on generic hardware and focus separately on localization, detection, or theoretical modeling.

\section{Conclusions and Future Work}
In this paper, we use the industry-recommended nRF5340 DK to develop a low-latency BLE-enabled PIN-based IoT authentication system for real-time data transfer. The performance of the proposed system is evaluated by measuring the RSSI with varying distances in four different scenarios. The common observation is that RSSI values decay logarithmically with respect to distance. Furthermore, the surroundings of the environment also influence RSSI values. 
In the future, the system can be improved by connecting more devices together via BLE mesh, allowing it to cover a larger area and work with multiple nodes. The RSSI part can also be improved by employing some basic filtering techniques to reduce the impact of noise and changing surroundings, resulting in more stable results. The system can be linked to a mobile app, allowing data to be viewed from anywhere rather than just the LCD. Security can also be improved by using encryption or another level of verification instead of just a PIN. Apart from that, more sensors can be added, allowing the same setup to be used for various types of monitoring in real-world scenarios.

\bibliographystyle{ieeetr}
\balance
\bibliography{references.bib}

\end{document}